\documentclass[12pt]{iopart}

\usepackage{iopams}
\usepackage{amsthm}
\usepackage{amstext}
\usepackage[utf8x]{inputenc}
\usepackage[backref]{hyperref}

\newenvironment{pr}{\begin{proof}[\textbf{Proof:}] \ }{\end{proof}}
\newtheorem{thm}{Theorem}[section]
\newtheorem{lem}[thm]{Lemma}
\newtheorem{prop}[thm]{Proposition}

\newtheorem{defi}[thm]{Definition}

\newcommand{\ep}{\epsilon}
\newcommand{\no}[2]{\|#1\|_{I_{#2},\infty}}
\newcommand{\noo}{\|F_1\|_{I_3,\infty}}
\newcommand{\Li}{\mathrm{Lip}}

\newcommand{\supp}{\mathrm{supp}}

\newcommand{\x}{x[x_0,\dot{x}_0]}
\newcommand{\y}{y[x_0,\dot{x}_0]}
\newcommand{\xe}{x_\ep[x_0,\dot{x}_0]}

\newcommand{\dx}{\dot{x}[x_0,\dot{x}_0]}
\newcommand{\dy}{\dot{y}[x_0,\dot{x}_0]}
\newcommand{\ve}{v_\ep[v_0,\dot{v}_0]}
\newcommand{\dve}{\dot{v}_\ep[v_0,\dot{v}_0]}
\newcommand{\vv}{v[v_0,\dot{v}_0]}
\newcommand{\Cinf}{\mathcal{C}^\infty}
\newcommand{\D}{\ensuremath{{\mathcal D}}}
\newcommand{\R}{\mathbb{R}}

\renewcommand{\d}{\ensuremath{\partial}}

\begin{document}
\title{On the completeness of impulsive gravitational wave space-times}
\author{Clemens S\"amann, Roland Steinbauer}
\address{Faculty of Mathematics, University of Vienna, Nordbergstra\ss e 15,
A-1090 Wien, Austria}
\eads{\mailto{clemens.saemann@univie.ac.at},
\mailto{roland.steinbauer@univie.ac.at}}


\begin{abstract}
We consider a class of impulsive gravitational wave space-times, which generalize 
impulsive pp-waves. They are of the form 
$M=N\times\mathbb{R}^2_1$, where $(N,h)$ is a  Riemannian manifold of arbitrary
dimension and $M$ carries the line element 
$ds^2=dh^2+ 2dudv+f(x)\delta(u)du^2$ with $dh^2$ the line element of $N$ and 
$\delta$ the Dirac measure. We prove a completeness result for such
space-times $M$ with complete Riemannian part $N$.
\end{abstract}

\section{Introduction}
Plane-fronted gravitational waves with parallel rays---pp-waves, for short---are
defined by the existence of a covariantly constant null vector field
$\mathbf{k}$ and are usually associated with the line element  in the so-called
Brinkmann form  
\begin{equation}\label{ppw}
  ds^2=2dudv+(dx^1)^2+(dx^2)^2+H(x^1,x^2,u)du^2
\end{equation}
on $\mathbb{R}^4$.
These space-times model gravitational or electromagnetic waves and other forms
of null matter and have been extensively studied (see e.g.\ \cite[Ch.\ 17]{GP09}
and the literature cited therein). The geodesic null congruence with tangent
$\mathbf{k}$ is non-expanding, shear-free, and twist-free and the latter
property implies the existence of a family of $2$-surfaces perpendicular to
$\mathbf{k}$ which are interpreted as wave surfaces. Moreover, since
$k^\mu_{\ ;\nu}$ vanishes, they are planar and rays orthogonal to them are
parallel.

It should be noted, however, that Brinkmann, who studied these geometries in the
context of conformal mappings of Einstein spaces (\cite{B25}), also included a
rotational term (rediscovered by Bonner (\cite{B70}) and recently studied further
under the name gyraton (\cite{F07})), as well as allowed for a general wave
surface. Including the latter effect, i.e., allowing for a Riemannian manifold
of arbitrary dimension as the wave surface we arrive at the following geometry
$(M,g)$: Let $(N,h)$ be a connected Riemannian manifold of dimension $n$, set
$M=N\times\mathbb{R}^2_1$ and equip $M$ with the line element
\begin{equation}\label{npw}
 ds^2\ =\ dh^2 + 2dudv + H(x,u)du^2, 
\end{equation}
where $dh^2$ denotes the line element of $(N,h)$. Moreover $u$, $v$ are global
null-coordinates on the $2$-dimensional Minkowski space $\mathbb{R}^2_1$ 
and $H:N\times\mathbb{R}\to\mathbb{R}$ is a smooth function.

These models have been studied in a series of papers by J. Flores and M. Sanchez
in part together with A. Candela (\cite{CFS03,FS03,CFS04,FS06}) mainly focusing
on causality and geodesics. These geometries allow one to shed some light on some of
the peculiar causal properties especially of plane waves (i.e., pp-waves
(\ref{ppw}) with $H(x^1,x^2,u)=h_{ij}(u)x^ix^j$), see e.g.\ \cite[Ch.\
13]{BEE96}. They turn out to be caused by the high degree of symmetries of plane 
waves and the fact that the wave surfaces of (\ref{ppw}) are flat $\mathbb{R}^2$. 

In \cite{CFS03} space-times of the form (\ref{npw}) have been called (general)
plane-fronted waves (PFW). However, by the geometric interpretation given above
and by the analogy with pp-waves it seems more natural to us to call the
space-times (\ref{npw}) \emph{$N$-fronted waves with parallel rays (NPW)}, which
we shall do from now on.

It turns out that the behaviour of $H$ at spatial infinity, i.e., for ``large
$x$'' is decisive for many of the global properties of NPWs. In order to
formulate precise statements we recall that one says that $H$ behaves
\emph{subquadratically at spatial infinity } if there exist a fixed point $\bar
x\in N$, continuous functions $0\leq R_1$, $R_2$ and a continuous function $p<2$
such that for all $(x,u)\in N\times\mathbb{R}$
\begin{equation}\label{sq} 
H(x,u)\leq R_1(u)d(x,\bar x)^{\,p(u)}+R_2(u). 
\end{equation} 
Here $d$ denotes the Riemannian distance function on $N$. Similarly we say that
$H$ behaves at most quadratically respectively superquadratically if $p\leq 2$ respectively 
$p>2$. In \cite{FS03} it has been shown that the causality of NPWs depends
crucially on the exponent $p$ in (\ref{sq}), with $p=2$ being the critical case.
In particular, NPWs are causal but not necessarily distinguishing, they are
strongly causal if $-H$ behaves at most quadratically at spatial infinity and
they are globally hyperbolic if $-H$ is subquadratic and $N$ is complete.
Similarly the global behaviour of geodesics in NPWs is governed by the behaviour
of $H$ at spatial infinity. From the explicit form of the geodesic equations it
follows (\cite[Thm.\ 3.2]{CFS03}) that a NPW is complete if and only if 
$N$ is complete and
\[
 D^{(N)}_{\dot x}\dot x=\frac{1}{2}\ \nabla_xH(x,s)
\]
has complete trajectories. Here $D^{(N)}_{\dot x}$ is the induced covariant
derivative on $N$ and $\nabla_x$ denotes the spatial gradient. Applying
classical results on complete vector fields (e.g.\ \cite[Thm.\ 3.7.15]{AMR88})
completeness of $M$ follows for autonomous $H$ (i.e., independent of $u$) in
case $H$ grows at most quadratic at spatial infinity. Clearly this implies
completeness for at most quadratic sandwich waves, that is, waves with $H$ compactly 
supported in $u$. 

In this work we consider \emph{impulsive }NPWs (INPWs), i.e., we set
$H(x,u)=f(x)\delta(u)$ in (\ref{npw}), where $\delta(u)$ is the Dirac measure on
the hypersurface $\{u=0\}$. Impulsive \emph{pp-waves } (for a summary see \cite[Ch.\
20]{GP09}) have been introduced by Penrose using a ``scissors-and-paste method''
(e.g.\ \cite{P72}) gluing two halves of Minkowski space along the null
hypersurface $\{u=0\}$ with a warp. On the other hand, impulsive pp-waves arise
as ultrarelativistic limits of Kerr-Newman black holes, the prototype being the
Aichelburg-Sexl geometry (\cite{AS71}).

The distributional term in the metric of impulsive pp-waves and INPWs makes it a
delicate matter to mathematically deal with these space-times; for a general
account on distributional geometries in GR see \cite{SV06}. Therefore impulsive
pp-waves have been treated using the nonlinear distributional geometry
(\cite{GKOS01}) built upon algebras of generalized functions (\cite{C85}). In
particular, geodesics in impulsive pp-wave space-times have been considered in
\cite{B97,S98}, and in \cite{KS99}, where an existence and uniqueness result for
the geodesic equations has been proved. From a global point of view these
results imply that impulsive pp-waves are geodesically complete. 

In this short note we prove a completeness result for INPWs with complete $N$.
We do so without using any theory of nonlinear distributions leaving a detailed
study of INPWs as distributional geometries to a subsequent paper. More
precisely, we view INPWs as geometries with a small but finitely extended
impulse: Let $\delta_\ep$ be some smooth approximation of the Dirac-delta 
(i.e., $\delta_\ep\to\delta$ weakly as $\ep\to 0$) and for fixed $\ep>0$
consider the metric 
\begin{equation}\label{inpw} 
ds_\ep^2\ =\ dh^2 + 2dudv + f(x)\delta_\ep(u)du^2 
\end{equation} 
on $M$, where $f$ is an arbitrary smooth function on $N$. We will show that for
any geodesic $\gamma$ in $(M,ds_\ep^2)$ there is $\ep_0$ small enough, such that $\gamma$
can be defined for all values of an affine parameter provided $\ep\leq\ep_0$.
Moreover the size of $\ep_0$ for which the geodesic becomes complete can be
explicitly estimated in terms of (derivatives of) $f$ and the initial data of
$\gamma$. Finally, we also show that the globally defined geodesics converge to
the geodesics of the background $N\times\mathbb{R}^2_1$ which, however, have to
be joined with a suitable warp at the shock hypersurface.

\section{The geodesic equations for INPWs}\label{sec:geo}
In this section we derive the geodesic equations for INPWs and fix some notation
to be used in the remainder of this work. We start by making precise the class
of regularizations we use for the Dirac delta. We set $I:=(0,1]$.
\begin{defi}\label{def-sdn}
A net $(\delta_\ep)_{\ep\in I}$ of smooth functions on $\mathbb{R}$ is called a
\emph{strict delta net} if it satisfies the following three properties.
\begin{enumerate}
  \item \label{def-sdn-L1b}
    The supports shrink to zero, $\supp(\delta_\ep)\to\{0\}$ for
    $\ep\searrow 0$.
  \item \label{def-sdn-conv}
    The integrals converge to $1$,  
    $\displaystyle\int_{\mathbb{R}}{\delta_\ep(x) \rmd x} \to 1$ for
    $\ep\searrow 0$.
  \item The $L^1$-norms are uniformly bounded, 
    $\displaystyle\exists K > 0:\ 
     \int_{\mathbb{R}}{|\delta_\ep(x)| \rmd x} \leq K\quad\forall \ep\in I$.
\end{enumerate}
\end{defi} 

Observe that this is a very general class of approximations of $\delta$. (Even
although smoothness excludes ``boxes", nets arbitrarily close to  ``boxes" and even
discontinuous regularizations are practically included by the fact that
$\Cinf_c$ is dense in $L^1$.) Without loss of generality we will always assume
that $\supp(\delta_\ep) \subseteq (-\ep,\ep)$ for all $\ep\in I$.

Now let $M=N\times \mathbb{R}^2_1$ be an INPW with $N$ a connected
$n$-dimensional and complete Riemannian manifold and let $M$ be endowed with the
family of line elements (\ref{inpw}), where $(\delta_\ep)_\ep$ is a strict delta net.

Denoting the Christoffel symbols of the Riemannian manifold $(N,h)$ by
$\Gamma^{(N)}$ one obtains the non-vanishing Christoffel symbols for $M$ with respect to 
a coordinate system $(x^1,\dots,x^n)$ of $N$ and $(u,v)$ null-coordinates of 
${\mathbb R}^2_1$

\begin{eqnarray*}
 \Gamma^k_{\,ij}&=&\Gamma^{(N)k}_{\hphantom{(N)}\,ij}
   \quad\mbox{for all $1\leq i,j,k\leq n$},\\
 \Gamma^{v}_{uj}&=&\Gamma^v_{ju}\,=\,
                   \frac{1}{2}\, \frac{\partial f}{\partial x^j}\, \delta_\ep
   \quad\mbox{for all $1\leq j\leq n$},\\
 \Gamma^v_{uu}&=&\frac{1}{2}\, f\, \dot\delta_\ep,\\
 \Gamma^k_{uu}&=&-\frac{1}{2}\, h^{km}\,
                 \frac{\partial f}{\partial x^m}\,\delta_\ep.
\end{eqnarray*}

Since all Christoffel symbols of the form $\Gamma^u_{jk}$ vanish we may use
$u$ as an affine parameter (thereby only excluding geodesics parallel to the 
shock hypersurface). Hence the geodesic equations reduce to the following set
of $n+1$ equations
\begin{eqnarray}\label{eq-geo}
 \ddot{v}_\ep 
  &=& -\frac{\partial f}{\partial x^j}(x_\ep)\ \dot{x}^j_\ep\ \delta_\ep 
      -\frac{1}{2}\ f(x_\ep)\ \dot{\delta_\ep}\label{eq-geo-v},\\
 D^{(N)}_{\dot{x}_\ep}\dot{x}_\ep 
  &=& \frac{1}{2}\ \nabla_xf(x_\ep)\ \delta_\ep.\label{eq-geo-x}\ 
\end{eqnarray}
Here $D^{(N)}$ and $\nabla_x$ denote the covariant derivative respectively the
gradient with respect to $h$. First observe that equation (\ref{eq-geo-v}) can be
integrated once the second equation has been solved. So we have to concentrate
on equation (\ref{eq-geo-x}), which is just the perturbed geodesic equation on
$N$ with potential $f$ and the non-autonomous term $\delta_\ep$. Moreover, since
the latter vanishes for $|u|\geq \varepsilon$ the $x$-component of the
geodesics on $M$ will for large $u$  coincide with the (unperturbed) geodesics
on $N$. By completeness of $N$ the question of completeness of $M$ reduces to
the question whether all perturbed geodesics on $N$ that enter the
regularization strip at  $u=-\varepsilon$ also leave it at $u=\varepsilon$, that
is whether the perturbed geodesics blow up before $u=\varepsilon$ or not.

Bearing this in mind we apply the following procedure to solve
the geodesic equation on $M$ as well as to address the problem of geodesic
completeness of $M$. We fix $\ep>0$ and impose initial data  $x_0\in N$,
$\dot{x}_0\in T_{x_0}N$ at $u=-1$ ``long before'' the shock and then follow the
unperturbed Riemannian geodesic on $N$ with this data, i.e., the solution of 
\[
 D_{\dot{x}}^{(N)}\dot{x}=0,\quad x(-1)=x_0,\quad \dot{x}(-1)=\dot{x}_0,
\]
which we denote by $\x$. By completeness of $N$ this geodesic $\x$ will reach
the shock region at $u=-\varepsilon$ and until then it will also be a solution
of the perturbed geodesic equation (\ref{eq-geo-x}) with the same data, which we
will denote by $\xe$. With this notation we have $\xe=\x$ on $]-\infty,-\ep]$
and to continue $\xe$ into the shock region $|u|\leq\ep$ we consider the initial
value problem 
\[
 \mbox{(\ref{eq-geo-x}) with data}\quad
 x_\ep(-\ep)=\x(-\ep), \quad \dot{x}_\ep(-\ep)=\dx(-\ep).
\]
To prove that $\xe$ extends to all values of the
parameter $u$ we only have to show that the latter initial value problem
possesses a solution denoted by $\tilde\xe$ until $u=\ep$, since for $u\geq\ep$
the right hand side of (\ref{eq-geo-x}) vanishes and we are solving the
(unperturbed) geodesic equation in the complete manifold $N$. That is, we only
have to show that no blow-up occurs within the shock region $|u|\leq\ep$,
which, in fact, will be done in the next section 
(at least for $\ep$ small enough). In total we will then have the global
perturbed geodesic $\xe$  
\begin{equation}\label{pgeo}
 \xe(u)=\left\{
         \begin{array}{rl}\x&u\leq-\ep\\
                          \tilde\xe &-\ep\leq u .
         \end{array}\right.
\end{equation}

Finally, as observed above, once we have such a solution of the $x$-component of
the geodesic in $M$ the equation for $v$ can be integrated to give a solution
for all $u\in{\mathbb R}$ and we will use the following notation: for initial
conditions $v_0, \dot{v}_0\in\mathbb{R}$ we denote by $\vv$ the straight line
$\vv(u)=v_0+\dot{v}_0(1+u)$, i.e., a solution of (\ref{eq-geo-v}) for $u\leq
-\ep$ and similarly $\ve$ denotes a solution of (\ref{eq-geo-v}) with
$\ve(-\ep)=\vv(-\ep)$ and $\dve(-\ep)=\dot{v}_0$.

\section{Completeness}

We now show that for any geodesic in $M$ we can choose $\ep$ sufficiently small
such that the geodesic can be extended through the shock. More precisely we
prove that (using the notation introduced above) the initial value problem 
\begin{equation}\label{ro_ivp}\fl
 D^{(N)}_{\dot{x}_\ep}\dot{x}_\ep=\frac{1}{2}\,\nabla_x f(x_\ep)\,\delta_\ep,
 \quad x_\ep(-\ep)=\x(-\ep),
 \quad \dot{x}_\ep(-\ep)=\dx(-\ep)
\end{equation}
has a local solution defined up to $u=\ep$, provided $\ep$ is small enough. 

\begin{prop}\label{prop-co}
For all $x_0\in N$, $\dot{x}_0\in T_{x_0}N$ there exists $\ep_0$ such that the
initial value problem (\ref{ro_ivp}) has a solution $\tilde\xe$ defined up to
$u=\ep$, provided $\ep\leq\ep_0$.
\end{prop}

The proof heavily rests on a fixed point argument which we provide in detail
in Lemma \ref{Lem-Basic-Ex} in the Appendix. Here we only observe that this
argument indeed provides the assertion of the Proposition.

\begin{pr}
We invoke Lemma \ref{Lem-Basic-Ex} with $b>0$, $c>0$,
$F_1(y,z)^k:=-{\Gamma_{ij}^{k(N)}(y)z^i z^j}$ (to express $D^{(N)}$ in local
coordinates) and $F_2(y)^k:=\frac{1}{2} {h^{km}(y)\frac{\d f}{\d{x^m}}(y)}$ which
is just $\frac{1}{2}\nabla_xf$ in coordinates. Clearly $F_1$ and $F_2$ are
smooth since $f$ and $h$ (and hence the Christoffel symbols) are assumed to be
smooth. Hence Lemma \ref{Lem-Basic-Ex} guarantees existence of a solution
$\tilde\xe$ of (\ref{ro_ivp}) until $u=\alpha-\ep$. So choosing
$\ep_0=\frac{\alpha}{2}$, the solution $\tilde\xe$ exists at least until
$u=\ep$, provided $\ep\leq\ep_0$.
\end{pr}

Observe that Lemma \ref{Lem-Basic-Ex} also implies that the solution $\tilde\xe$
together with its first derivative ${\dot{\tilde x}}_\ep[x_0,\dot x_0]$ is
uniformly bounded (in $\ep$ on $[-\ep_0,\ep_0]$). Moreover (\ref{alpha}) gives an
upper bound on $\ep_0$ in terms of the initial velocity $\dot x_0$ and of the
Christoffel symbols on $N$ as well as of $\nabla_x f$ on a neighborhood of the
data $x_0$. Next we state our completeness result.

\begin{thm}\label{thm-co}
For all $x_0\in N$, $\dot{x}_0\in T_{x_0}N$ and all $v_0$, $\dot v_0\in{\mathbb
R}$ there exists $\ep_0$ such that the solution $(\xe,\ve)$ of the geodesic
equation (\ref{eq-geo-v},\ref{eq-geo-x}) with initial data $x_\ep(-1)=x_0$,
$\dot x_\ep(-1)=\dot x_0$, $v_\ep(-1)=v_0$, $\dot v_\ep(-1)=\dot v_0$ is
defined for all $u\in{\mathbb R}$, provided $\ep\leq\ep_0$.
\end{thm}

\begin{pr}
Given $x_0$, $\dot x_0$, Proposition \ref{prop-co} provides us with $\ep_0$ such
that the solution of (\ref{ro_ivp}) is defined for $u\in[-\ep,\ep]$ for all
$\ep\leq\ep_0$. In this case we hence may define  $\xe$ as in (\ref{pgeo}) for
all $u\in{\mathbb R}$ so it only remains to integrate (\ref{eq-geo-v}) twice to
obtain a globally defined solution $v_\ep$. Hence in total we obtain a unique
globally defined geodesic ${\mathbb R}\ni u\mapsto (\xe(u),v_\ep[v_0,\dot
v_0](u))$.
\end{pr}

We point out that $\alpha$ in the proof of Proposition \ref{prop-co} and hence
$\ep_0$ for which the geodesic is defined on all of ${\mathbb R}$ depends on the
choice of the initial data $x_0$ and $\dot{x}_0$. Hence we can not, in general,
obtain a global bound $\ep_0$ such that for fixed $\ep\leq\ep_0$ the manifold
$M$ is geodesically complete. There are, however, two special cases where we
actually obtain geodesic completeness of $M$ for $\ep$ sufficiently small. First
assume that $N$ is compact. Then we obtain a globally defined $\ep_0$ since 
$x_0$ varies in a compact set only and upon reparametrization we may achieve
that $|\dot{x}_0|=1$. On the other hand, if $-f$ behaves subquadratically (cf.\
(\ref{sq})) then by the compactness of the support of $\delta_\ep$ we may apply
the results of \cite{FS03} mentioned in the introduction to obtain completeness
without even the need to invoke the fixed point argument.

However, one may say that ``in the limit $\ep\to 0$'' we obtain a geodesically
complete manifold, hence one may say that INPWs are geodesically complete
irrespectively of the behaviour of the profile function $f$. This is in sharp
contrast to the case of extended NPWs where completeness depends crucially on
the behavior of $H$ at ``spatial infinity'': the role of the
$x$-asymptotics of $H$ becomes irrelevant in the impulsive limit. 

However, the precise meaning of the completeness statement (i.e., the
dependence of $\ep_0$ on the data) is encoded in the formulation of our theorem
above. A more straight forward completeness result for INPWs can be provided
using nonlinear distributional geometry (\cite{GKOS01,KS02}) in the sense of
J.F. Colombeau (\cite{C85}), and we will address this topic in a subsequent
paper.

\section{Limits}
In this section we compute the limits of the global geodesics derived above as
$\ep\to 0$. We start by analyzing the $x$-component and introduce some more
notation in the same spirit as at the end of section \ref{sec:geo}. We define
the prospective limit of $\xe$ by pasting together the solution $\x$ of
the unperturbed equation for $u<0$ with an appropriate solution of the
unperturbed equation for $u>0$. To this end denote by $\tilde\x$ the solution of
the (unperturbed) geodesic equation on $N$ with data $\tilde x(0)=\x(0)$ and
$\dot{\tilde x}(0)=\dx(0)+ \frac{1}{2}\nabla_xf(\x(0))$. Finally denote the
prospective limit by
\begin{equation}\label{ro:xlim}
 \y(u):=\left\{\begin{array}{ll}\x(u)&u\leq 0\\\tilde\x(u)&u\geq
0.\end{array}\right.
\end{equation}
Observe that $\y$ is a continuous curve ${\mathbb R}\to N$ which is
piece-wise smooth with a single break point at $u=0$. Moreover it is not
differentiable (in general) as we have
\begin{eqnarray*}
  \lim_{u\nearrow0}\dy(u)&=&\dx(0),\\
  \lim_{u\searrow0}\dy(u)&=&\dx(0)+\frac{1}{2}\nabla_xf(\x(0)).
\end{eqnarray*}

For simplicity we write $F_1(y,z)^k:=-{\Gamma_{ij}^{k(N)}(y)z^i z^j}$ and
$F_2^k=\frac{1}{2}\nabla^k_x f=\frac{1}{2}h^{kl}\frac{\partial f}{\partial x^l}$ 
as in the proof of Theorem \ref{prop-co} and start with an auxiliary result needed 
throughout the remainder of this section.

\begin{lem} \label{Lem-unif-conv}
The global solution $\xe$ of (\ref{eq-geo-x}) (defined in (\ref{pgeo}))
satisfies 
\[
 \xe(\ep u) \to \y(0) = \x(0) \ \mbox{uniformly on $[-1,1]$ as $\ep\searrow
0$.}
\]
\end{lem}

\begin{pr}
To keep the notation transparent we abbreviate $\x$ by $x$ and
$\xe$ by $x_\ep$. We have
\begin{equation*}\fl\qquad
\sup_{u\in[-1,1]}|x_\ep(\ep u)-x(0)|\leq \sup_{u\in[-1,1]}\underbrace{|x_\ep(\ep
u)-x(\ep u)|}_{=:(\star)} +
\sup_{u\in[-1,1]}|x(\ep u)-x(0)|\ .
\end{equation*}
The second term goes to zero as $\ep\searrow0$ since $x$ is uniformly continuous
on compact sets. To estimate the first term
we integrate the differential equations for $x_\ep$ and $x$ (see also (\ref{intop}) 
in the Appendix) to obtain
\begin{eqnarray*}\fl
 (\star)
 &\leq& \int\limits_{-\ep}^{\ep u} \int\limits_{-\ep}^s
    |F_1(x_\ep(r),\dot{x}_\ep(r))-F_1(x(r),\dot{x}(r))|
    \rmd r\rmd s 
    + \int\limits_{-\ep}^{\ep u} \int\limits_{-\ep}^s |F_2(x_\ep(r))|
    |\delta_\ep(r)|\rmd r\rmd s
 \\ \fl
 &\leq&
 C_1 \ep^2 +  C_2\|\delta_\ep\|_{L^1}\ep 
 \ \leq\  C \ep \ \to\  0\ (\ep \searrow 0)\ ,
\end{eqnarray*}
where we have used that by Lemma \ref{Lem-Basic-Ex}, $x_\ep$ and $\dot{x}_\ep$
are bounded independently of $\ep$ and the constants $C_1$ and $C_2$ contain the
$L^\infty$-norms of $F_1$ and $F_2$ respectively on suitable compact sets.
\end{pr}


\begin{prop}\label{Prop-xe-lim}
The global solution $\xe$ of (\ref{eq-geo-x}) (defined in (\ref{pgeo}))
satisfies
\begin{eqnarray*}
 \xe&\to&\y\quad
  \mbox{uniformly on compact subsets of $\R$},\\
 \dot\xe&\to& \dot\y\quad
  \mbox{uniformly on compact subsets of $\R\backslash\{0\}$}.
\end{eqnarray*}
\end{prop}

\begin{pr}
Again we write $x$ for $\x$ and $x_\ep$ for $\xe$ and similarly $y$ for $\y$ and
$\tilde x_\ep$ for $\tilde\xe$. Without loss of generality we only consider the
compact interval $[-1,1]$. We distinguish three cases: $-1\leq u \leq -\ep$,
$-\ep \leq u \leq \ep$ and $\ep\leq u\leq 1$. 

In the first case $x_\ep=x=y$ on $[-1,-\ep]$ (and hence $\dot{x}_\ep = \dot{x}$
on the same interval), since $x_\ep$ and $x$ solve the same initial value
problem. If $-\ep\leq u\leq\ep$ the result for $x_\ep$ follows immediately from
Lemma~\ref{Lem-unif-conv} while for the derivative $\dot x_\ep$ there is nothing
to prove in this case.

Finally, for $\ep\leq u \leq 1$ we observe that $x_\ep=\tilde x_\ep$ and
$y=\tilde{x}$ solve the same differential equation but now with different
initial conditions, namely $\tilde x_\ep(\ep),\ \dot{\tilde x}_\ep(\ep)$, and
$\tilde x(\ep)$ and $\dot{\tilde x}(\ep)$, respectively. By continuous
dependence on the initial data we obtain
\begin{equation*}
 \max(|\tilde x_\ep(u)-\tilde{x}(u)|,
      |\dot{\tilde x}_\ep(u)-\dot{\tilde{x}}(u)|)
 \leq \max(|\tilde x_\ep(\ep)-\tilde x(\ep)|,
           |\dot{\tilde x}_\ep(\ep)-\dot{\tilde{x}}(\ep)|)\ e^{L},
\end{equation*}
where $L$ is a Lipschitz constant of $F_1$ on the compact image of $[0,1]$
under $\tilde x$, $\dot{\tilde x}$, $\tilde x_\ep$, $\dot{\tilde x}_\ep$, and it
suffices to estimate the difference of the data. Indeed we have
\[
 |\tilde x_\ep(\ep)-\tilde x(\ep)|\ \leq\  |\tilde x_\ep(\ep)-\tilde x(0)| \ 
 +\ |\tilde x(0)-\tilde x(\ep)| \ \to\ 0,
\]
since the first term converges to zero by Lemma~\ref{Lem-unif-conv} and the
second by continuity. Similarly we have
\[
 |\dot{\tilde x}_\ep(\ep)-\dot{\tilde x}(\ep)|\ \leq\  
  |\dot{\tilde x}_\ep(\ep)-\dot{\tilde x}(0)| \ +\ |\dot{\tilde
x}(0)-\dot{\tilde x}(\ep)|,
\]
where again the second term on the right hand side goes to zero by continuity.
To estimate the first term we plug in the integral representation of
$\dot{\tilde x}_\ep$ to obtain
\begin{eqnarray*}
 |\dot{\tilde x}_\ep(\ep)-\dot{\tilde x}(0)| 
 &=&
 |\dot{\tilde x}_\ep(\ep)-\dot x(0)-F_2(x(0))|\\
 &\leq&
 |\dot{\tilde x}_\ep(-\ep)-\dot x(0)|+ \int\limits_{-\ep}^\ep|F_1(\tilde
x_\ep(s),\dot{\tilde x}_\ep(s))|ds\\
 &&\hphantom{|\dot{\tilde x}_\ep(-\ep)-\dot x(0)|}
 +\left| \int\limits_{-\ep}^\ep F_2(\tilde x_\ep(s))\, \delta_\ep(s)\,ds-
 F_2(x(0))\right|.
\end{eqnarray*}
Now the first term on the right hand side vanishes in the limit since 
$\dot{\tilde x}_\ep(-\ep)=\dot x(-\ep)\to\dot x(0)$. The second term goes
to zero again by the uniform boundedness of $\tilde x_\ep$ and $\dot{\tilde
x}_\ep$. To obtain the same conclusion for the third term we again take into
account the uniform boundedness of $\tilde x_\ep$ and the fact that
$(\delta_\ep)_\ep$ is a strict delta net.
\end{pr}

Next we turn to the $v$-component and recall that $(u,v)\in\R^2_1$ and so we may
work distributionally. 
\begin{prop}
The global solution $\ve$ of (\ref{eq-geo-v}) satisfies
\begin{eqnarray*}\fl
 \ve&\to&\vv -\frac{1}{2}\ f(x(0))H 
             -\Big(\dot{x}^j(0)+\frac{1}{4}\nabla_x f^j(x(0))\Big)
               D_j f(x(0))u_+,
\end{eqnarray*}
where $u_{+}(u)=uH(u)$ denotes the so-called kink function and we again have abbreviated $\x$ by $x$.
\end{prop}

\begin{pr}
In addition to the abbreviations $x$ and $x_\ep$ used already above we write $v$
for $\vv$ and $v_\ep$ for $\ve$. Since we have $v_\ep(u)=v_0+\dot{v}_0\cdot(1+u) +
H*H*\ddot{v}_\ep(u)$ and since convolution is a separately continuous
operation, it suffices to calculate the distributional limit of $\ddot{v}_\ep$.
Inserting the integral representation of $\dot{x}^j_\ep$ into equation
(\ref{eq-geo-v}) we obtain
\begin{eqnarray*}\label{eq-ddotv}\fl
 \ddot{v}_\ep(u)
 =-\underbrace{D_jf(x_\ep(u))\,\delta_\ep(u)\,\dot{x}_\ep^j(-\ep)}_{(I)} 
     -\frac{1}{2}\underbrace{D_jf(x_\ep(u))\,\delta_\ep(u)\,
                  \int\limits_{-\ep}^{u}\!F_1^j(x_\ep(s),\dot{x}_\ep(s))\,ds}_{(II)}\\
  \fl\hphantom{\ddot{v}_\ep(u)=}
  -\frac{1}{2}\underbrace{D_jf(x_\ep(u))\,\delta_\ep(u)\,
                  \int\limits_{-\ep}^{u}\!F_2^j(x_\ep(s))\delta_\ep(s)\,ds}_{(III)} 
  -\frac{1}{2}\underbrace{
                  (f(x_\ep(u))\dot\delta_\ep(u))}_{(IV)}.
\end{eqnarray*}
It is easily seen that $(I)\to\dot{x}^j(0)D_j f(x(0))\delta$ 
and that $(IV)\to f(x(0))\dot{\delta}$ in $\D'(\mathbb{R})$.
On the other hand $(II)\to 0$ in $\D'(\R)$ since for all test functions
$\phi\in\D(\R)$ we have (again using the uniform boundedness of $x_\ep$ and
$\dot x_\ep$) 
\[
 \left|\int\limits_{\R}\phi(u)\,D_j f(x_\ep(u))\,\delta_\ep(u)
  \int\limits_{-\ep}^u{F_1^j(x_\ep(r),\dot{x}_\ep(r))\rmd r}\rmd u\right| 
 \leq 2\ep\,C\,\|\phi\|_\infty.
\]
Finally, we show that $(III)\to \frac{1}{2}D_j
f(x(0))F_2^j(x(0))\delta$. Indeed, we have
\begin{eqnarray*}
 \fl{\left|\int\limits_\R{\phi(u)D_j
  f(x_\ep(u))\,\delta_\ep(u) \int\limits_{-\ep}^u{F_2^j(x_\ep(r))\delta_\ep(r)\rmd
r}\rmd u}
  -\frac{1}{2}\phi(0)
  D_j f(x(0))F^j_2(x(0))\right|} \\
 \fl\quad\leq \left|\int\limits_{-\ep}^\ep {\phi(u)\delta_\ep(u)
   \int\limits_{-\ep}^u{F^j_2(x_\ep(r))\delta_\ep(r)\rmd r}
   \Big(D_j f(x_\ep(u))-D_j f(x(0))\Big)\rmd u}\right|\\
 \fl\qquad +
  \left |\int\limits_{-\ep}^\ep {\phi(u)\delta_\ep(u)
  \int\limits_{-\ep}^u{\Big(F^j_2(x_\ep(r))-F^j_2(x(0))\Big)\delta_\ep(r)\rmd r}
  \rmd u}\right| |D_j f(x(0))|\\
 \fl\qquad +
  \left|\int\limits_{-\ep}^\ep{\phi(u)\delta_\ep(u)\int\limits_{-\ep}^u{\delta_\ep(r)
  \rmd r}\rmd u
  -\frac{1}{2}\phi(0)}\right||D_j f(x(0))||F^j_2(x(0))| \\
 \fl\quad\leq C\, \|\phi\|_\infty
   \sup_{u\in[-1,1]}{\Big|D_j f(x_\ep(\ep u)-D_jf(x(0))\Big|}\\
 \fl\qquad + C\, \|\phi\|_\infty
   \sup_{u\in[-1,1]}{\Big|F_2^j(x_\ep(\ep u))-F^j_2(x(0))\Big|} \\
 \fl\qquad + C\,
   \left|\int\limits_{-\ep}^\ep{\phi(u)\delta_\ep(u)\int\limits_{-\ep}^u{\delta_\ep(r)
   \rmd r}\rmd u
   -\frac{1}{2}\phi(0)}\right|,
\end{eqnarray*}
where we have absorbed all constant terms into the ``generic constant'' $C$.
Now the first and the second term converge to zero, again by Lemma
\ref{Lem-unif-conv}. Finally, the integral term in the last line converges to
zero by an elementary calculation.
\end{pr}

Summing up we have shown that the $x$-component of the
limit is continuous but has a kink at the shock hypersurface. The $v$-component,
however, is not even continuous but has a jump at the shock in
addition to a kink. The parameters of the kinks and the jump are given in terms
of the profile function $f$ and its derivatives at the point where the geodesic
hits the shock hypersurface. So globally the geodesics on $M$ are given by
geodesics on the background $N\times\R^2_1$, which have to be joined suitably at
the shock hypersurface.

This result complements the completeness result (Theorem \ref{thm-co})) of section 3:
the globally defined geodesics in the complete limiting space-time are given by suitably gluing
together the geodesics of the background space-time at the shock hypersurface.

\ack{This work was partially supported by projects Y237, P20525,
and P23714 of the Austrian Science Fund.}

\appendix
\section*{Appendix}
\addcontentsline{toc}{section}{Appendix}
\setcounter{section}{1}
\setcounter{thm}{0}
\def\thesection{\Alph{section}}

In this appendix we detail the fixed point argument used in the proof of our
main result. Our argument is built on a slightly sharper version of the
Banach fixed point theorem (see \cite{W52}). Indeed the integral operator
$A_\ep$ used below to solve the initial value problem is \emph{not} a contraction
on the naturally chosen Banach space $X_\ep$ and so the Banach fixed
point theorem does not apply.
\begin{thm}(Weissinger's fixed point theorem)
Let $X$ be a nonempty closed subset of a Banach space $(E,\|.\|)$. Moreover let
$\sum_{n=1}^\infty{a_n}$ be a convergent
series of positive real numbers $(a_n)_n$ and $A:X\rightarrow X$ a map with the
property that
\begin{equation}
\|A^n(u)-A^n(v)\|\leq a_n \|u-v\|\quad \forall u,v\in X\ \forall
n\in\mathbb{N}.
\end{equation}
Then $A$ has a unique fixed point. 
\end{thm}

Now we state and prove our main technical result. For brevity we write $\|{F}\|_{I,\infty}$
for the $L^\infty$-norm of the function $F$ on the set $I$.

\begin{lem}\label{Lem-Basic-Ex}
Let $F_1\in \mathcal{C}^{\infty}(\mathbb{R}^{2n},\mathbb{R}^n)$, $F_2 \in
\mathcal{C}^{\infty}(\mathbb{R}^n,\mathbb{R}^n)$,
let $x_0,\dot{x}_0\in\mathbb{R}^n$, let $b>0$, $c>0$ be given and let
$(\delta_\ep)_\ep$ be a strict delta net 
with $L^1$-bound $K>0$. Define $I_1:=\{x\in\mathbb{R}^n: |x-x_0|\leq b \}$,
$I_2:=\{x\in\mathbb{R}^n:|x-\dot{x}_0|\leq c +
K \no{F_2}{1} \}$ and $I_3:=I_1\times I_2$. Moreover set 
\begin{equation}\label{alpha}
\alpha:=\min\left(1,\ \frac{b}{|\dot{x}_0| + \noo+K \no{F_2}{1}}\,,\ \frac{c}{\noo}\right).
\end{equation}
Then the initial value problem
\begin{equation}\label{eq-geo1}
\begin{array}{ll}
\ddot x_\ep=F_1(x_\ep,\dot{x_\ep})+F_2(x_\ep)\delta_\epsilon, \\
x_\ep(-\epsilon)=x_0,\ \dot{x}_\ep(-\epsilon)=\dot{x}_0,
\end{array}
\end{equation}
has a unique solution $x_\ep$ on $J_{\epsilon}:=[-\epsilon,\alpha-\epsilon]$
with $(x_\ep(J_\ep),\dot x_\ep(J_\ep))\subseteq I_3$.
In particular, both $x_\ep$ and $\dot{x}_\ep$ are
bounded, uniformly in $\ep$.
\end{lem}

\begin{pr}
We consider the closed subset
$X_\epsilon:=\{x_\ep\in\mathcal{C}^{\infty}(J_\ep,\mathbb{R}^n):
x_\ep(J_\ep)\subseteq I_1, \dot{x}_\ep(J_\ep)\subseteq I_2\}$ of the Banach
space ${\mathcal C}^1(J_\ep,\R^n)$ with norm $\|x \|_{\mathcal
{C}^1}=\|x\|_{J_\ep,\infty}+\|\dot x\|_{J_\ep,\infty}$. We define the operator
$A_\ep$ on  $X_\ep$ by ($t\in J_\ep$)
\begin{equation}\label{intop}\fl
A_\ep(x_\ep)(t):=x_0 + \dot{x}_0\cdot(t\!+\!\ep) + \int\limits_{-\ep}^t\int\limits_{-\ep}^s\!
F_1(x_\ep(r),\dot{x}_\ep(r))\rmd r
\rmd s + \int\limits_{-\ep}^t\int\limits_{-\ep}^s\! F_2(x_\ep(r))\delta_\ep(r)\rmd r \rmd s.
\end{equation}
First we show that the operator $A_\ep$ maps $X_\ep$ to itself. Let
$x_\ep\in X_\ep$ and 
$t\in J_\ep$, then we have for the zero-order derivative of $A_\ep(x_\ep)$
\begin{eqnarray*}\fl
  \lefteqn{|A_\ep(x_\ep)(t)-x_0|}\\
 &\leq& |\dot{x}_0|(t+\ep) + \int\limits_{-\ep}^t\int\limits_{-\ep}^s\!
  |F_1(x_\ep(r),\dot{x}_\ep(r))|\rmd r \rmd s
  +\int\limits_{-\ep}^t\int\limits_{-\ep}^s\! |F_2(x_\ep(r))| |\delta_\ep(r)| \rmd r \rmd s\\
 &\leq& 
  \alpha\, |\dot{x}_0|+ \alpha^2\, \noo +
  \alpha\, \no{F_2}{1}\|\delta_\ep\|_{L^1}\\
 &\leq&
  \alpha\, (|\dot{x}_0| + \noo + K\no{F_2}{1})\ \leq \ b,
\end{eqnarray*}
and for the first-order derivative 
\begin{eqnarray*}\fl
 |\case{\rmd}{\rmd t}(A_\ep(x_\ep))(t)-\dot{x}_0|
&\leq& \int\limits_{-\ep}^t\! |F_1(x_\ep(r),\dot{x}_\ep(r))|\rmd r + \int\limits_{-\ep}^t\!
 |F_2(x_\ep(r))| |\delta_\ep(r)| \rmd r \\ 
&\leq&
 \alpha\, \no{F_1}{3} + \no{F_2}{1}\|\delta_\ep\|_{L^1}\ \leq\ c +
 K \no{F_2}{1}.
\end{eqnarray*}
At this point we claim that we can find a sequence of positive real numbers
$(a_n)_{n\geq 2}$ such that $\sum_{n=2}^\infty{a_n}<\infty$
and
$$
\|A_\ep^n(x_\ep)-A_\ep^n(y_\ep)\|_{\mathcal{C}^1(J_\ep)}\leq a_n\,
\|x_\ep-y_\ep\|_{\mathcal{C}^1(J_\ep)}.
$$
So let $\mathbb{N}\ni n\geq 2$, $x_\ep,y_\ep\in X_\ep$, and
$t\in J_\ep$. Denoting by $[n\int\limits_{-\ep}^t]$ the $n$-times iterated
integral we obtain (with $\Li(F_i,I_j)$ a Lipschitz constant for $F_i$ on $I_j$)
\begin{eqnarray*}\fl
 \lefteqn{|A_\ep^n(x_\ep)(t)-A_\ep^n(y_\ep)(t)|}\\
 &\leq& [2n\int\limits_{-\ep}^t]|F_1(x_\ep(r),\dot{x}_\ep(r))-F_1(y_\ep(r),
        \dot{y} _\ep(r))|\rmd^{2n}r\\
 &&+
  {[2n\int\limits_{-\ep}^t]}|F_2(x_\ep(r))-F_2(y_\ep(r))| |\delta_\ep(r)|\rmd^{2n}r\\
 &\leq&
 \Big(\Li(F_1, I_3)\ [2n\int\limits_{-\ep}^t]\rmd^{2n}r\\
 &&+ 
\Li(F_2,I_1)\|\delta_\ep\|_{L^1}\ [(2n\!-\!1)\int\limits_{-\ep}^t]\rmd^{2n\!-\!1}r\Big)
\|x_\ep-y_\ep\|_{\mathcal{C}^1(J_\ep)}\\
 &\leq&
 \Big(\Li(F_1, I_3)\frac{\alpha^{2n}}{(2n)!} +
\Li(F_2,I_1)\|\delta_\ep\|_{L^1}\ \frac{\alpha^{2n-1}}{(2n-1)!}\Big)
\|x_\ep-y_\ep\|_{\mathcal{C}^1(J_\ep)}\ .
\end{eqnarray*}
Furthermore for the derivative of $A_\ep^n$ we find that
\begin{eqnarray*}\fl
 \lefteqn{|\case{\rmd}{\rmd t}(A_\ep^n(x_\ep))(t)-\case{\rmd}{\rmd
  t}(A_\ep^n(y_\ep))(t)|}\\
 &\leq&
  {[(2n\!-\!1)\int\limits_{-\ep}^t]}|F_1(x_\ep(r),\dot{x}_\ep(r))-F_1(y_\ep(r),
\dot{y}_\ep(r))|\rmd^{2n-1}r\\
 &&+ {[(2n\!-\!1)\int\limits_{-\ep}^t]}|F_2(x_\ep(r))-F_2(y_\ep(r))|
 |\delta_\ep(r)|\rmd^{2n-1}r\\
 &\leq&\Big(\Li(F_1, I_3)\ [(2n\!-\!1)\int\limits_{-\ep}^t]\rmd^{2n-1}r\quad \\
 &&\qquad\qquad+ \Li(F_2,I_1)\|\delta_\ep\|_{L^1}\
  [(2n\!-\!2)\int\limits_{-\ep}^t]\rmd^{2n-2}r\Big)
  \|x_\ep-y_\ep\|_{\mathcal{C}^1(J_\ep)}\\
 &\leq&
  \Big(\Li(F_1, I_3)\frac{\alpha^{2n-1}}{(2n-1)!} +
  \Li(F_2,I_1)\|\delta_\ep\|_{L^1}\ \frac{\alpha^{2n-2}}{(2n-2)!}\Big)
\|x_\ep-y_\ep\|_{\mathcal{C}^1(J_\ep)}.
\end{eqnarray*}
Summing up we obtain
\begin{eqnarray*}\fl
\|A_\ep^n(x_\ep)-A_\ep^n(y_\ep)\|_{\mathcal{C}^1(J_\ep)}\leq 4 \max\Big(\Li(F_1,
I_3),K\,\Li(F_2,I_1)\Big)\ \frac{\alpha^{2n-2}}{(2n-2)!}\ 
\|x_\ep-y_\ep\|_{\mathcal{C}^1(J_\ep)},
\end{eqnarray*}
which proves our claim. 

Now Weissinger's fixed point theorem provides us with
the existence of a unique solution $x_\ep\in X_\ep$. 

Finally, since $x_\ep,\ \dot{x}_\ep$ stay in $I_1$ respectively $I_2$ (which are
defined independently of $\ep$), $x_\ep$ and $\dot{x}_\ep$ are bounded by $b$ and
$c+K \no{F_2}{1}$, respectively.
\end{pr}

\end{document}